\documentclass{raa}
\usepackage{anysize}
\usepackage{natbib}
\usepackage[latin2]{inputenc}
\usepackage{amsmath}
\usepackage{fixmath}
\usepackage{wrapfig}
\usepackage{graphicx,times}
\usepackage{epstopdf}
\usepackage{hyphenat}
\usepackage{gensymb}

\begin{document}

\title{Analysis of the Instability Growth Rate During \\the Jet -- Background Interaction in the Magnetic Field}
 \volnopage{Vol.0 (200x) No.0, 000--000}      
   \setcounter{page}{1}          

   \author{M. Hork\'y \inst{1,2}
\and P. Kulh\'anek\inst{1}}

   \institute{Czech Technical University in Prague, Faculty of Electrical Engineering, Department of Physics, \mbox{Technick\'a~2, 166~27 Prague, Czech Republic}; {\it horkymi1@fel.cvut.cz}\\
    \and
             Astronomical Institute of the Academy of Sciences of the Czech Republic,  \mbox{Bo\v{c}n\'i~II 1401/1a,  141~31 Prague 4, Czech Republic}
   }

   \date{Received~~2012 October 30}

\abstract{The two-stream instability is a common instability, responsible for many observed phenomena in nature, especially the interaction of jets of various origin with the background plasma (e.g. extragalactic jet interacting with the cosmic background). The dispersion relation not considering magnetic field is described by the well-known Buneman relation. In 2011, Bohata, B\v re\v n, and Kulh\'anek derived the relation for the two-stream instability without the cold limit, with the general orientation of magnetic field, and arbitrary stream directions. The maximum value of the imaginary part of the individual dispersion branches $\omega_{n}(k)$ is of interest from the physical point of view. It represents the instability growth rate which is responsible for the turbulence mode onset and subsequent reconnection on the ion radius scale accompanied by a strong plasma thermalization. The paper presented here is focused on the non-relativistic instability growth rate and its dependence on various input parameters, such as magnitude and direction of magnetic field, sound velocity, plasma frequency of the jet and direction of the wave vector during the jet -- background interaction. The results are presented in well-arranged plots and can be used for determination of the plasma parameter values close to which the strong energy transfer and thermalization between the jet and the background plasma occur.
}

   \authorrunning{M. Hork\'y, P. Kulh\'anek}            
   \titlerunning{Instability Growth Rate During the Jet -- Background Interaction}  

   \maketitle

The most common plasma instabilities are the two-stream instabilities, which can occur during a plasma jet interaction with the plasma background. Such situations are observed in astrophysical processes, e.g. interaction of the galactic jets with the intergalactic medium \cite{Silk2012} or interaction of star jets with the interstellar medium \cite{Murphy2008}. Oskar Buneman derived the basic dispersion relation describing such instabilities in the late 1950's for cold unmagnetized plasmas \cite{Buneman1959}. The magnetohydrodynamic instabilities in the ideal plasma are discussed in \cite{Bonanno2011}. Magnetic fields are crucial for the phenomena taking place in jets \cite{Urpin2006}.  In 2011, Bohata et al. published a paper containing the derivation of the non-relativistic dispersion relation for magnetized plasmas and without the cold limit restriction \cite{Bohata2011}. It is called it the Generalized Buneman Dispersion Relation (GBDR) and it is described by the equation

\begin{equation}
\begin{gathered}
\prod_{\alpha=1}^2  \left\{ \Omega_{\alpha}^4 - \Omega_{\alpha}^2 \left[{\rm i} \frac {  {\bf F}_{\alpha}^{(0)} \cdot {\bf k}} {m_{\alpha} }+c_{{\rm s}\alpha}^2 k^2+\omega_{{ \rm p}\alpha}^2 + \omega_{ {\rm c}\alpha}^2 \right]
- \frac {\Omega_{\alpha} \omega_{ {\rm c}\alpha}} {m_{\alpha}}\left ({\bf F}_{\alpha}^{(0)} \times { \bf k}\right) \cdot {\bf e}_{\rm B} \right.\\
+~\left. \omega_{ {\rm c}\alpha}^2 ({ \bf k} \cdot {\bf e}_{\rm B}) \left[{\rm i} \frac {  {\bf F}_{\alpha}^{(0)} \cdot {\bf e}_{\rm B}} {m_{\alpha} }+\left(c_{{\rm s}\alpha}^2 k^2+\omega_{{ \rm p}\alpha}^2\right) \frac {{ \bf k} \cdot {\bf e}_{\rm B}}{k^2}\right]\right\}
- \prod_{\alpha=1}^2 \frac {\omega_{{ \rm p}\alpha}^2}{k^2}\left [ \Omega_{\alpha}^2 k^2 -  \omega_{ {\rm c}\alpha}^2  \left({\bf e}_{\rm B} \cdot { \bf k}\right)^2\right] = 0,
\end{gathered}
\end{equation}

\noindent where $\Omega_{\alpha}= \omega - {\bf k} \cdot {\bf u}_{\alpha}^{(0)}$ is the Doppler shifted frequency, $\omega_{ {\rm c}\alpha}$ is the cyclotron frequency, $\omega_{ {\rm p}\alpha}$ is the plasma frequency, $ {\bf F}_{\alpha}^{(0)}$ is the Lorentz force, ${\bf e}_{\rm B}$ is the unit vector in the direction of the magnetic field and $c_{{\rm s}\alpha}$ is the sound velocity. Index  $\alpha$ denotes the corresponding media (jet or background).

In the previous work \cite{Bohata2011}, the numerical solution for the case of two identical plasma beams with the same velocities, but opposite directions was found for various input parameters. The situation of the plasma jet penetration into the plasma background was studied as well and the numerical solution for this problem was found \cite{poster,sppt}.

The maximum of the imaginary part of the solution is denoted as the Plasma Instability Growth Rate (PIGR). This paper is focused on finding the plasma parameters for which this maximum occurs (the non-relativistic case and the plasma jet interaction with the plasma background are assumed). The calculations are performed on the microscopic level and using the linear approximation. For the plasma parameters leading to the maximum of the imaginary part of the dispersion relation, the instability arises and amplitudes of all variables grow exponentially. In such a situation, the linear approximation is no longer valid, and other methods for modelling of the physical phenomena must be introduced. One of the possibilities is the Particle in Cell (PIC) simulation, \cite{Stockem2008}. The results of these calculations can be therefore applied to: 1)~The search for the instability regimes in which strong thermalization, turbulence, micro-reconnections on ion radius, non-thermal radiation, shock onset and other interesting phenomena can occur. The subsequent PIC simulations of the plasma behaviour leading to significant phenomena seems to be the most reasonable next step of the research in this regime. 2)~The tests of the acceptance of the PIC codes (the PIC code must lead to an instability onset for the parameters calculated by the method proposed in next paragraph).

\section {Method}
Rhe following indices were designated in our analysis: ``j'' for parameters of the jet and ``b'' for parameters of the background. It is beneficial to transform the variables and the whole GBDR relation to the dimensionless form. After this step the relations simply scalable and the equations are covariant against to this transformation. It suggest that the results can be used for both space and laboratory plasmas, such as  thermalization in astrophysical jets or in fusion experiments. The relations for the dimensionless form were chosen with regard to the zero background velocity as \cite{poster,sppt}:
\begin{eqnarray}
\label{nondim}
&\overline{c}_{\rm s j} \equiv \frac {c_{\rm s j}}{u_{\rm j}}\,, \hspace{0.3cm} \overline{c}_{\rm s b} \equiv \frac {c_{\rm s b}}{u_{\rm j}}\,,\nonumber\\
&\overline{\omega}_{\rm c j} \equiv \frac {\omega_{\rm c j}}{\omega_{\rm p b}}\,, \hspace{0.3cm} \overline{\omega}_{\rm c b} \equiv \frac {\omega_{\rm c b}}{\omega_{\rm p b}}\,,\nonumber\\
&\overline{\omega}_{\rm p j} \equiv \frac {\omega_{\rm p j}}{\omega_{\rm p b}}\,, \hspace{0.3cm} \overline{\omega}_{\rm p b} \equiv \frac {\omega_{\rm p b}}{\omega_{\rm p b}} = 1\,,\nonumber\\
&\overline{u}_{\rm b} \equiv \frac {u_{\rm b}}{u_{\rm j}}\,, \hspace{0.3cm} \overline{u}_{\rm j} \equiv \frac {u_{\rm j}}{u_{\rm j}} = 1\,,\\
&\overline{k} \equiv \frac {k u_{\rm j}}{\omega_{\rm p b}}\,, \hspace{0.3cm} \overline{\omega} \equiv \frac {\omega}{\omega_{\rm p b}}\,,\nonumber\\
&\overline{\Omega}_{\rm j} = \overline{\omega} - \overline{k} \cos \varphi \sin \theta_{\rm k}\,,\nonumber\\
&\overline{\Omega}_{\rm b} = \overline{\omega} - \overline{k} \overline{u}_{2} \cos \varphi \sin \theta_{\rm k}\,.\nonumber
\end{eqnarray}

\begin{figure}[!ht]
\centering
\includegraphics[width=6.75cm]{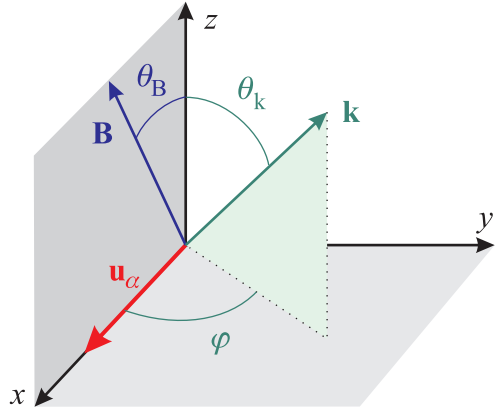}
\caption{The system of coordinates used in calculations.}
\label{Fig1}
\end{figure}

\noindent The reference system was set according to the Fig. \ref{Fig1}, in which the directions of the respective vectors ${\bf u}_{\alpha}$, $\bf B$ and $\bf k$ are drawn. The wave-vector can point in any direction, the magnetic field vector lies in the (\textit{x-z}) plane and the jet is directed along the \textit{x}-axis. The vector coordinates are
\begin{eqnarray}
&{\bf u}_{\alpha} = (u_{\alpha}, 0, 0)\,,\nonumber\\
&{\bf B} = (B \sin \theta_{\rm B}, 0,B \cos \theta_{\rm B})\,,\\
&{\bf k} = (k \cos \varphi \sin \theta_{\rm k}, k \sin \varphi \sin \theta_{\rm k}, k \cos \theta_{\rm k})\,.\nonumber
\end{eqnarray}

\noindent After simple manipulation the dimensionless form of the dispersion relation becomes \cite{poster,sppt}

\begin{eqnarray}
&\left[\overline{\Omega}_{\rm j}^4 + {\rm i} \overline{\Omega}_{\rm j}^2 \overline{\omega}_{{\rm cj}} \overline{k} (G_1) -  \overline{\Omega}_{\rm j}^2 \left(\overline{c}_{\rm sj}^2 \overline{k}^2+\overline{\omega}_{\rm pj}^2\right) - \overline{\Omega}_{\rm j}^2 \overline{\omega}_{{\rm cj}}^2 \right.\nonumber\\
&- \left. \overline{\Omega}_{\rm j}  \overline{\omega}_{{\rm cj}}^2 \overline{k} (G_3) + \left(\overline{\omega}_{{\rm cj}}^2 \overline{k}^2 \overline{c}_{\rm sj}^2 + \overline{\omega}_{{\rm cj}}^2 \overline{\omega}_{\rm pj}^2\right)(G_2)^2\right]\nonumber\\
&\cdot \left[\overline{\Omega}_{\rm b}^4 + {\rm i} \overline{\Omega}_{\rm b}^2 \overline{\omega}_{{\rm cb}} \overline{k} \overline{u}_{\rm b} (G_1) -  \overline{\Omega}_{\rm b}^2 \left(\overline{c}_{\rm sb}^2 \overline{k}^2 + 1\right) - \overline{\Omega}_{\rm b}^2 \overline{\omega}_{{\rm cb}}^2\right. \\
&- \left. \overline{\Omega}_{\rm b}  \overline{\omega}_{{\rm cb}}^2 \overline{k} \overline{u}_{\rm b} (G_3)+\left (\overline{\omega}_{{\rm cb}}^2 \overline{k}^2 \overline{c}_{\rm sb}^2 + \overline{\omega}_{{\rm cb}}^2\right )(G_2)^2\right]\nonumber\\
&- \left[\overline{\omega}_{\rm pj}^2 \left(\overline{\Omega}_{\rm j}^2 - \overline{\omega}_{{\rm cj}}^2 (G_2)^2\right)\right] \cdot \left[\overline{\Omega}_{\rm b}^2 - \overline{\omega}_{{\rm cb}}^2 (G_2)^2\right] = 0\,,\nonumber
\end{eqnarray}

\noindent where the goniometrical terms were denoted as
\begin{eqnarray}
&G_1 = (\cos \theta_{\rm B} \sin \varphi \sin \theta_{\rm k})\,,\nonumber\\
&G_2 = (\cos \varphi \sin \theta_{\rm k} \sin \theta_{\rm B} + \cos \theta_{\rm k} \cos \theta_{\rm B})\,,\\
&G_3 = (\cos^2 \theta_{\rm B} \cos \varphi \sin \theta_{\rm k} - \cos \theta_{\rm B} \cos \theta_{\rm k} \sin \theta_{\rm B})\,.\nonumber
\end{eqnarray}

\begin{figure}[!ht]
\begin{center}
\includegraphics[width=110mm]{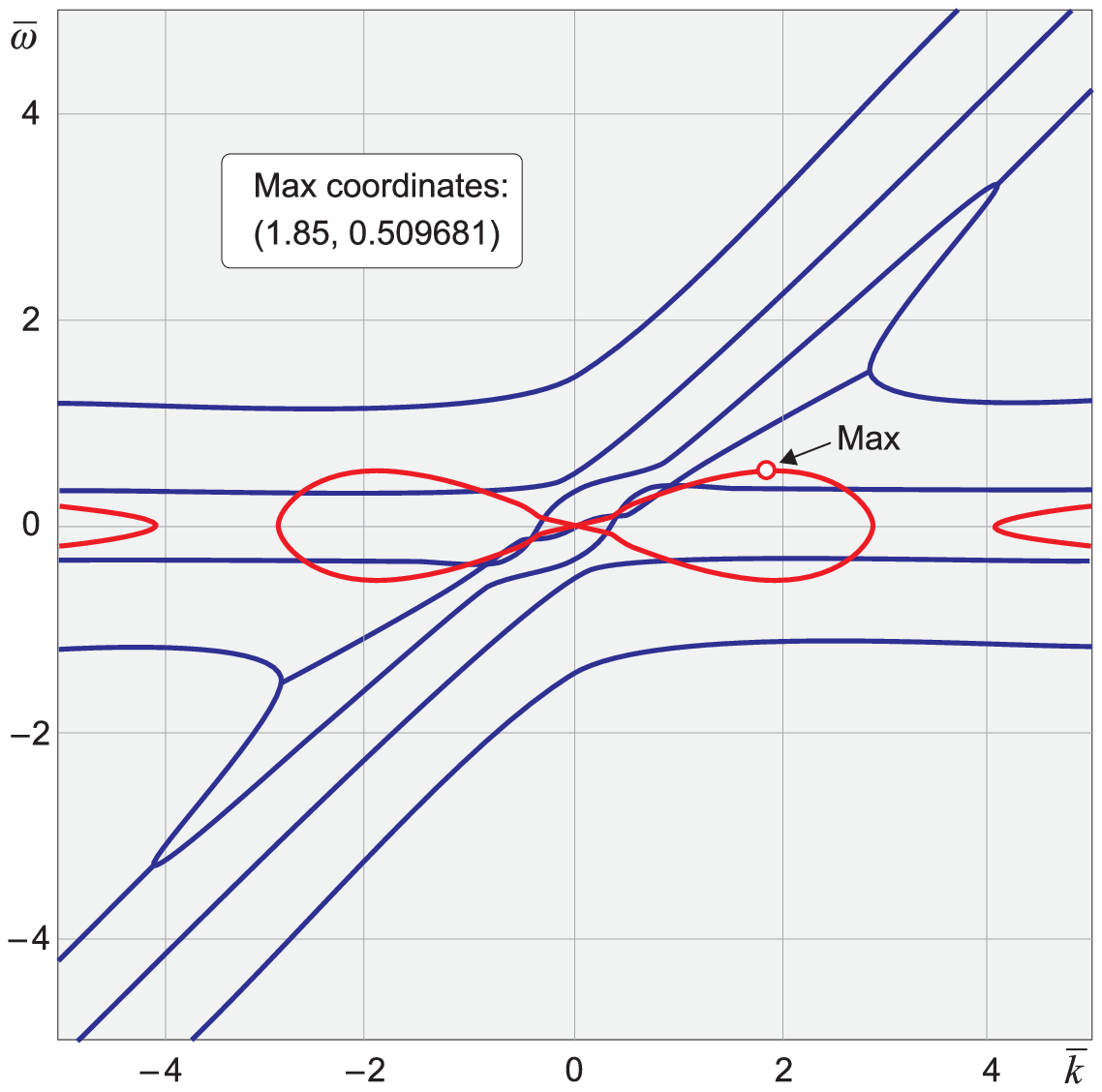}
\caption{Real (blue) and imaginary (red) branches of the GBDR dispersion relation and PIGR value (denoted as Max) for
\mbox{$\overline{\omega}_{{\rm cj}} = \overline{\omega}_{{\rm cb}} = 0.5$},
\mbox{$\overline{c}_{{\rm sj}} = \overline{c}_{{\rm sb}} = 0.1$},
\mbox{$\overline{\omega}_{{\rm pj}} = 1$},
\mbox{$\theta_{{\rm k}} = \pi / 2$},
\mbox{$\varphi = 0$},
\mbox{$\theta_{{\rm B}} = \pi / 4$}. These values were used as initial values for the calculations, see Table~1 for details.}
\label{Fig2}
\end{center}
\end{figure}

\noindent  This dimensionless relation is a polynomial equation with complex roots of the $8^{\rm{th}}$ order. The algorithm developed by Hubbard, Shleicher, and Sutherland \cite{Hubbard} was used to find the solution. The algorithm was implemented in the Wolfram Mathematica 8.0.1. and in its principle ensures that the iteration from the initial seeds will converge tol the roots in comparison with much more simpler Newton-Raphson method. The results were arranged into plots in which the real branches of the solution have different colour than the imaginary branches, and the maximum imaginary value determining the PIGR value was highlighted. The example of the program output is shown in the Fig. \ref{Fig2}.

In the next step, the PIGR value dependence on various parameters of the dimensionless GBDR (such as cyclotron frequencies of the jet and the background, the sound velocities of the jet and the background, the plasma frequency of the jet and the directions of the magnetic field and of the wave-vector) was found.

\section {Results}

The PIGR value was calculated during the program cycle running from the minimum to the maximum value of the tracked parameter while other parameters were fixed at their initial values. Intervals of these parameters are shown in Tab. \ref{tab1}. It was not necessary to change the jet velocity, because its dimensionless value was fixed at 1.

\renewcommand*\arraystretch{1.5}
\begin{table}[!ht]
\bigskip
\centerline{\begin{tabular}{|c|c|c|c|}
\hline
\ \ \ Parameter\ \ \
& \ \ \ Initial value\ \ \
& \ \ \ Minimum value\ \ \
& \ \ \ Maximum value\ \ \   \\
\hline
$\overline{\omega}_{{\rm cj}}\,,\ \overline{\omega}_{{\rm cb}}$
&$0.5$
&$0.1$
&$3.0$\\
\hline
$\overline{c}_{{\rm sj}}\,,\ \overline{c}_{{\rm sb}}$
&$0.1$
&$0.1$
&$1.5$\\
\hline
$\overline{\omega}_{{\rm pj}}$
&$1$
&$1$
&$5$\\
\hline
$\theta_{{\rm k}}$
&$\pi / 2$
&$0$
&$\pi / 2$\\
\hline
$\varphi$
&$0$
&$0$
&$\pi / 2$\\
\hline
$\theta_{{\rm B}}$
&$\pi / 4$
&$0$
&$\pi / 2$\\
\hline
\end{tabular}}
\caption{Parameters used for the numerical solution. The dispersion relation for the initial values is depicted on the Fig.~1.}
\label{tab1}

\end{table}

\subsection{The PIGR value dependence on the cyclotron frequencies}

\begin{figure}[ht]
\centerline{\includegraphics[width=110mm]{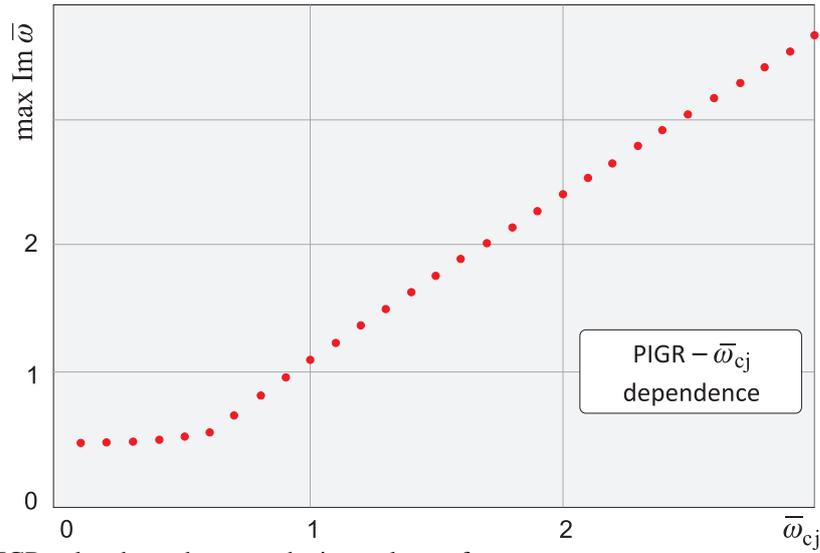}}
\vspace{-3mm}
\caption{The PIGR value dependence on the jet cyclotron frequency}
\label{Fig3}
\end{figure}

\begin{figure}[!htb]
\vspace{7mm}
\centerline{\includegraphics[width=110mm]{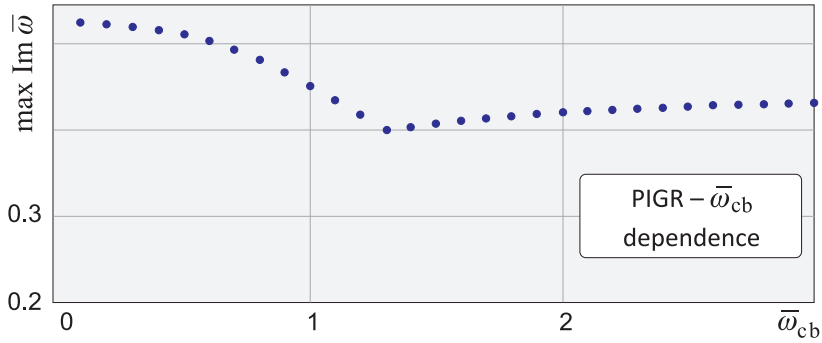}}
\vspace{-3mm}
\caption{The PIGR value dependence on the background cyclotron frequency}
\label{Fig4}
\end{figure}

The cyclotron frequency of the jet and the cyclotron frequency of the background were increased from the minimum value of $0.1$ to the final value of $3.0$ with the step size $0.1$. The cyclotron frequency is proportional to the magnetic field intensity influencing the charged particles. The PIGR value dependence on the cyclotron frequency of the jet is depicted in the Fig. \ref{Fig3},  where the almost almost linearly increasing character of this relation for $\overline{\omega}_{{\rm cj}}>0.6$ is noticeable. The change of the slope at this point ($\overline{\omega}_{{\rm cj}}=0.6$) corresponds to the localization of the minimum of the two different imaginary branches of the dispersion relation. The PIGR value dependence on the cyclotron frequency of the background is more complicated than the case of jet cyclotron frequency. In the Fig. \ref{Fig4}, the descending character for lower frequency values is visible.  The curve reaches minimum and then it rises to an asymptote. The  minimum was numerically determined to be $\overline{\omega}_{{\rm c}b}=1.313$ and the corresponding PIGR value was equal to $0.39795$. This effect is caused by the fact that the solution has two imaginary branches in this area and while the value of $\overline{\omega}_{{\rm c}b}$ is increasing the first branch is descending and the second is rising. In the minimum both branches have equal PIGR values.

\subsection{The PIGR value dependence on the sound velocities}

\begin{figure}[!ht]
\vspace{7mm}
\centerline{\includegraphics[width=110mm]{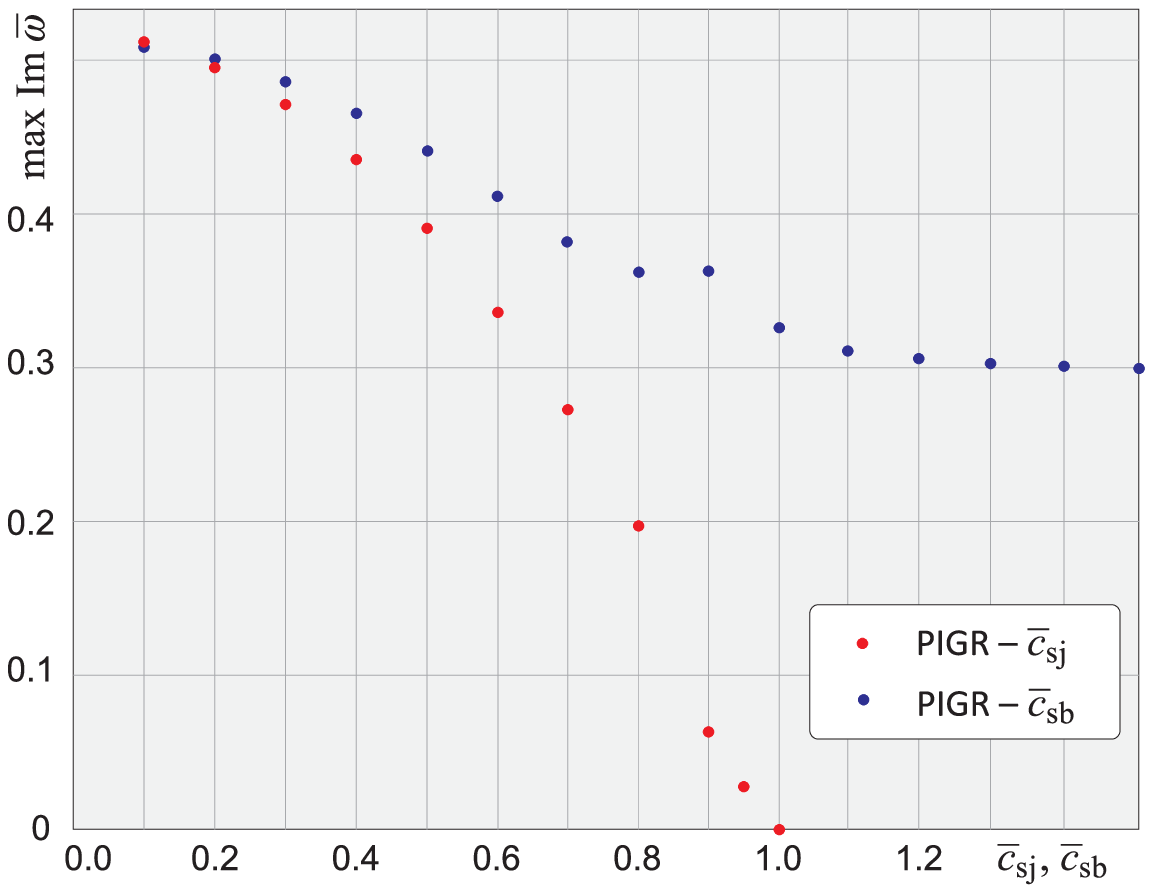}}
\vspace{-3mm}
\caption{The PIGR dependence on the sound velocities}
\label{Fig5}
\end{figure}

\begin{figure}[!ht]
\centerline{\includegraphics[width=110mm]{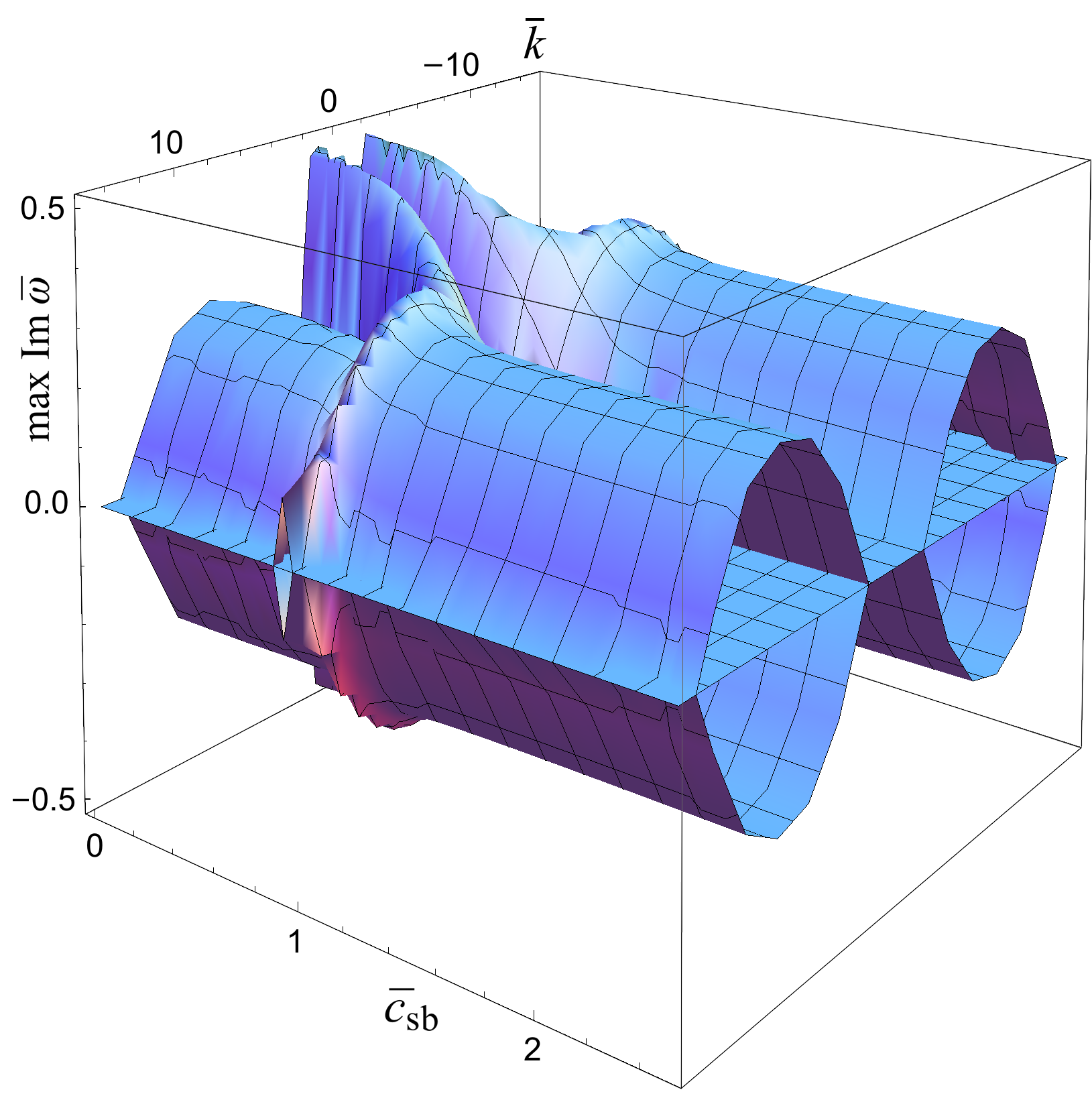}}
\caption{The imaginary branches dependence on $\overline{c}_{{\rm sb}}$}
\label{Fig6}
\end{figure}

Sound velocity is proportional to $(T_\alpha/m_\alpha)^{1/2}$, where $T_\alpha$ is the plasma temperature, and $m_\alpha$ is the mass of the jet or of the background particles (electrons or ions). The index $\alpha$ labels the corresponding media (jet or background). Modification of the original Buneman dispersion relation by addition of the sound velocities of both media is a~result of the calculation with non-zero pressure, i.e. without the cold limit.  The dimensionless parameter $\overline{c}_{{\rm s}}$ involves the plasma jet velocity, see Eq.~(\ref{nondim}), and $\overline{c}_{{\rm sj}} > 1$ indicates subsonic jet and vice versa. Both the sound velocity of the jet and the sound velocity of the background were increased from the initial value $0.1$ to the final value  $1.5$ with the step $0.1$. The PIGR value dependence on the sound velocities of both jet and background is depicted in the Fig.~\ref{Fig5}. The jet dependence (red circles) has descending character and PIGR value is zero, while $\overline{c}_{{\rm sj}} \geq 1$. It implies that for a subsonic jet (in dimensionless form the sound velocity equals 1) the GBDR relation has no imaginary branch and therefore the PIGR value is zero and no instabilities occur. The PIGR value dependence on the sound velocity of the background (blue circles) is more complicated. The interesting peak resides at the value $\overline{c}_{{\rm sb}} \doteq 0.9$. We made three-dimensional plot of the imaginary branches of the GBDR solution to uncover the~origin of this local maximum. The result can be seen in the Fig.~\ref{Fig6}. First axis corresponds to $\overline{c}_{{\rm sb}}$, second to $\overline{k}$, and the vertical axis to the value of the imaginary branch of the PIGR coefficient. It is clear now that the peak originates from the ridge present at the dispersion relation solution.

\subsection{The PIGR value dependence on the plasma frequency of the jet}
\begin{figure}[!ht]
\centerline{\includegraphics[width=110mm]{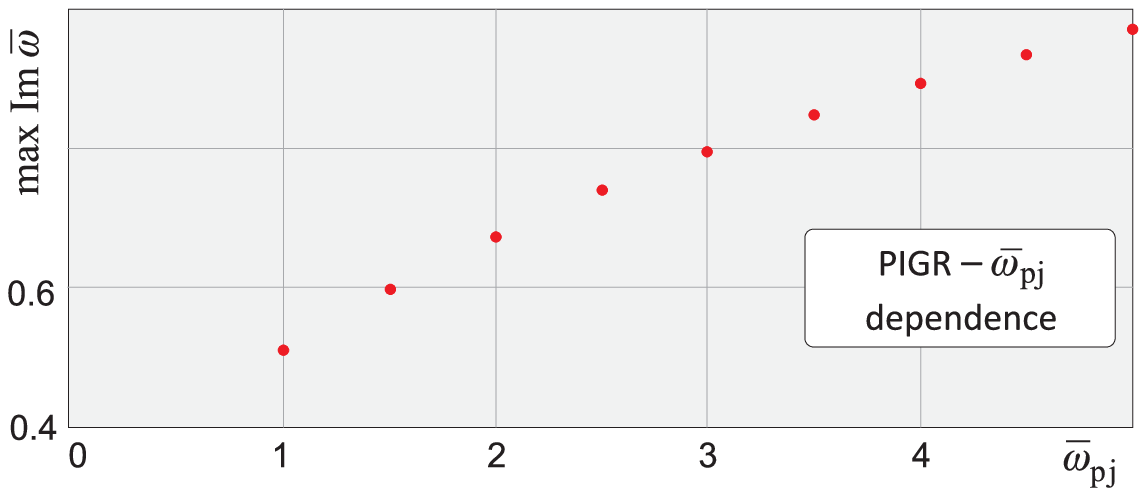}}
\caption{The PIGR value dependence on $\overline{\omega}_{{\rm pj}}$}
\label{Fig7}
\end{figure}

All dimensionless frequencies in the system are related to the background plasma frequency, see Eq.~(\ref{nondim}). It means that the dimensionless plasma frequency of the background $\overline{\omega}_{{\rm pb}}$ is equal by definition 1 and dimensionless plasma frequency of the jet $\overline{\omega}_{{\rm pj}}$ is in fact the ratio of the jet and the background plasma frequencies. This parameter is therefore proportional to the $(n_{\rm ej}/n_{\rm eb})^{1/2}$. During numerical calculation it was increased from the initial value $1$ to the final value $5$ with the step size $0.5$. It is a rather big step, but as it can be seen i then Fig.~\ref{Fig7}, the dependence is very simple and  without any discontinuities or local maxima or minima.

\subsection{The directional PIGR value dependencies}
\begin{figure}[!ht]
\centerline{\includegraphics[width=110mm]{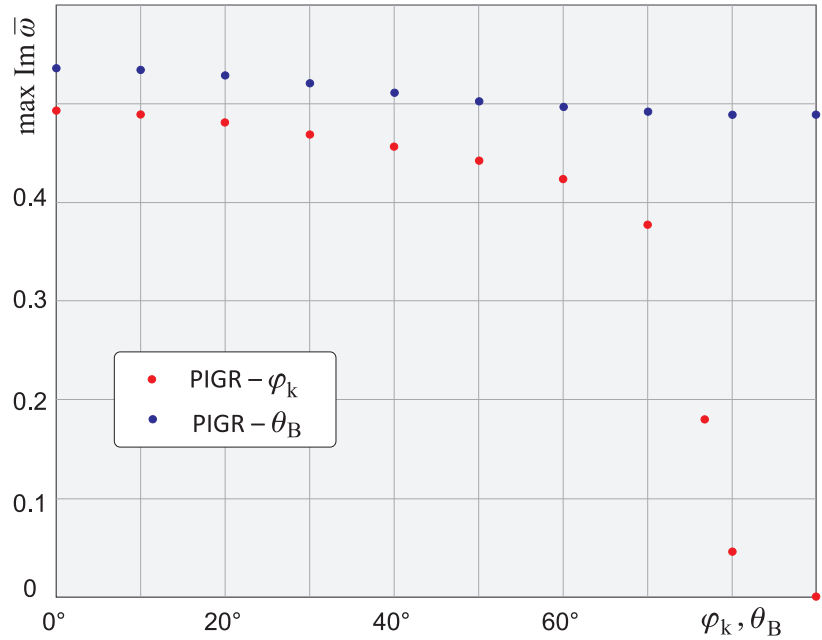}}
\caption{The PIGR dependence on the magnetic field direction (blue circles) and on the  wave vector direction (red circles)}
\label{Fig8}
\end{figure}

The PIGR value dependence on the magnetic field direction is simply predictable from the Lorentz equation of motion. Longitudinal magnetic field will evoke less disturbances than the perpendicular one. As you can see in the Fig.~\ref{Fig8}, the PIGR value has maximum at $\theta_{\rm B} = 0$ (perpendicular direction) and decreases for increasing $\theta_{\rm B}$. The PIGR value dependence on the direction of the wave vector is also predictable due to the dot product between $\bf{k}$ and ${\bf{u}}_{\alpha}$ in the GBDR relation, so if the angle between the wave vector and the velocity equals $90 \degree$, the PIGR value should be zero. In the Fig.~\ref{Fig8}, the dependence has descending character and it's zero at the angle $90 \degree$. Because of the cylindrical symmetry, both angles $\varphi_{\rm k}$ and $\theta_{\rm{B}}$ were altered only from $0 \degree$ to $90 \degree$ with the step of $10 \degree$.

\section{Conclusion}

Plasma jets from black holes and other types of astronomical objects are driven by the magnetic fields, and classical Buneman instability analysis (without magnetic fields) is inapplicable. All calculations must be performed using the Generalized Buneman Dispersion Relation (GBDR) with nonzero pressure and nonzero magnetic field. The Plasma Instability Growth Rate (PIGR) as the maximum of the imaginary parts of the GBDR relation was numerically calculated in this paper. The PIGR value is responsible for strong thermalization during the jet-background interaction and these calculations can be useful for understanding of the underlying processes. Furthermore, the known PIGR value can be used as a simple test of Particle in Cell (PIC) numerical methods frequently used for the plasma jet simulations. It is an interesting but still open question if the PIGR value could be calculated analytically directly from the dimensionless GBDR relation. The dispersion relation is not anisotropic in the velocity space. This possibility can cause other phenomena, e.g. particle acceleration, shock origin, etc. \cite{Nishikawa2006,Mizuno2009}, which will be the topic of detailed PIC simulations.

\subsection*{Acknowledgements}
\emph{Research described in this paper was supported by the Czech Technical University in Prague grants SGS10/266/OHK3/3T/13 (Electric discharges, basic research and application), SGS12/181/OHK3/3T/13 (Plasma instabilities and plasma-particle interactions) and by the Grant Agency of the Czech Republic grant GD205/09/H033 (General relativity and its applications in astrophysics and cosmology).}

\makeatletter
\renewcommand\@biblabel[1]{}
\makeatother

\end{document}